# Investigating drug translational research using PubMed articles

Xin Li[1] and Xuli Tang[2]

[1] xl60@hust.edu.cn
School of Medicine and Health Management, Tongji Medical College, Huazhong University of Science and Technology, Wuhan (China)

[2] xltang@ccnu.edu.cn
School of Information Management, Central China Normal University, Wuhan (China)

**Abstract**
Drug research and development are embracing translational research for its potential to increase the number of drugs successfully brought to clinical applications. Using the publicly available PubMed database, we sought to describe the status of drug translational research, the distribution of translational lags for all drugs as well as the collaborations between basic science and clinical science in drug research. For each drug, an indicator called Translational Lag was proposed to quantify the interval time from its first PubMed article to its first clinical article. Meanwhile, the triangle of biomedicine was also used to visualize the status and multidisciplinary collaboration of drug translational research. The results showed that only 18.1% (24,410) of drugs/compounds had been successfully entering clinical research. It averagely took 14.38 years (interquartile range, 4 to 21 years) for a drug from the initial basic discovery to its first clinical research. In addition, the results also revealed that, in drug research, there was rare cooperation between basic science and clinical science, which were more inclined to cooperate within disciplines.

**Introduction**

Despite the ongoing progress in science and technology, there has been a delay in translating a newly discovered compound/drug into a cure for human diseases and difficulty in successfully getting it approved by the U.S. Food and Drug Administration (FDA). Only around 10% of compounds/ drugs that entered the phase 1 clinical trials are finally approved, and it usually costs more than $2.5 billion for every drug approved (K. Lee et al., 2019; Williams et al., 2015). What's more, almost 99% of compounds/drugs died before entering into clinical trials, which is known as "the valley of death" (Li et al., 2020). Hence, translation science, as defined as the bridge between basic research and clinical research, has recently surged in the field of drug research and development.

Much of the recent studies in translational research have been focused on identifying the key factors that inhibit or promote the basic-to-clinical translation of basic discovery. For example, Hutchins et al. (2019) proposed a machine learning system to identify the pathways of knowledge flow from research outputs to clinical applications and pointed out that distinct knowledge pathways are significantly associated with the success of translation. K. Lee et al. (2019) tracked more than 2.4 thousand human genes along the translational continuum and concluded that the choice of research object for basic research may influence the further translation to clinical applications because of its translational value. Meanwhile, several studies have reviewed translational research in the field of drug research and development. These reviews mainly analyzed and summarized the definition, process, and methodologies of drug translational research, often illustrated with an instance of successful translation, such as Zidovudine for HIV infection, Levamisole for colon cancer, and Tamoxifen for breast cancer prevention (Li et al., 2020). However, the extant studies failed to describe the drug translational research from a bird's eye view, and also lacked quantitative aspects of research.

Herein, we conduct a bibliometric analysis of drug translational research to investigate the status of drug translational research, the distribution of translational lags for all drugs, and the basic-clinical collaborations in drug research. We define translational lag for a drug as the time interval between the publication time of its earliest basic article and the publication time of its earliest clinical article. Our results show that only 18.1% (24,410) of drugs/compounds have been successfully entering into the phase of clinical research. It averagely takes 14.38 years (interquartile range, 4 to 21 years) for a drug from the initial basic discovery to its first clinical research. We also find that, in drug research, there was rare cooperation between basic science and clinical science, which are more inclined to cooperate within disciplines.

**Methodology**

*Data and pre-processing*

Our analysis is based on the PubMed dataset. With around 30 million biomedical citations, PubMed is considered one of the best-curated sources of biomedical literature and has been widely used in both academia and industry. We downloaded the PubMed 2020 baseline in XML files from its official website. The total number of PubMed articles was 30,477,134, in which over 29M articles were published from 1965 to 2018. An XML parser was developed using JAVA and dom4j, to extract the metadata of the 29M PubMed articles, such as titles, abstracts, MeSH terms, publication date, and journals. We then adopted BioBERT (J. Lee et al., 2019), which was a pre-trained biomedical language model based on PubMed and PubMed Central (PMC), to extract biomedical entities (for example, diseases, drugs/compounds, genes/proteins, and mutations) from titles and abstracts of PubMed articles. BioBERT has been considered as the state-of-art for the recognition of biomedical entities, and has been successfully employed for multiple biomedical tasks, such as knowledge graph building and text mining (Li et al., 2020; J. Lee et al., 2019; Xu et al., 2020). With transformers and self-attention, BioBERT have advantages over the previous methods that are based on rules, conditional random fields or long short-term memory, such as the PKDE4J (Song et al., 2015) and the PubTator (Wei et al., 2019). Next, we filtered those articles that didn't mention any drugs/compounds in their titles and abstracts. Finally, 9,093,936 articles mentioning 134,561 drugs/compounds (published between 1965-2018) with their metadata were stored in a local MySQL database for further analysis.

*Article classification and visualization*

To track drugs along the translational continuum using PubMed articles, it is crucial to classify articles into basic research and clinical research. In the light of weber's algorithm, articles were classified as Animal (A), Molecule/Cell (C), Human (H), or a combination of these three (including AC, AH, CH, ACH) using MeSH terms assigned to each article (Weber, 2013). Specifically, terms with MeSH codes beginning with subtree code A11, B02, B03, B04, and G02.111.570 are Molecule/Cell- related MeSH terms, i.e., C terms; terms with MeSH codes starting with subtree code B01.050.150.900.649.313.988.400.112.400.400 or M01 are Human-related MeSH terms, i.e., H terms; and terms with MeSH codes beginning with subtree code B01, except B01.050.150.900.649.313.988.400.112.400.400 are Animal-related MeSH terms, i.e., A terms. Each article could have one or more of these three types of MeSH terms, or none of the three (Hutchins et al., 2019; Weber, 2013; Li et al., 2021). Note that the version of the Medical Subject Headings (MeSH) that we used in this study is the MeSH 2020.

We also employed the triangle of medicine to visualize the translational drug research, in which the three-dimension (A, C, and H) scores into an X-Y coordinate system (Hutchins et al., 2019; Li et al., 2020). The specific coordinate for an article is defined as follows ($N_A$, $N_C$ and $N_H$ represent the number of A, C and H terms in an article).

$$X = N_A \times \frac{\sqrt{3}}{2} + N_C \times (-)\frac{\sqrt{3}}{2} + N_H \times 0 \qquad (1)$$
$$Y = N_A \times (-)\frac{1}{2} + N_C \times (-)\frac{1}{2} + N_H \times 1 \qquad (2)$$

Specifically, each article was plotted on a triangle with three vertices: A $(\frac{\sqrt{3}}{2}, -0.5)$, C $(-\frac{\sqrt{3}}{2}, -0.5)$ and H $(0, 1)$. Different articles were represented by a circle filled with different colors, and the size of the circle can reflect the number of articles sharing the same coordinates.

*Drug translational research analysis*

The whole procedure of drug translational research generally involves two major parts: (1) basic research with the new drug design and test in animals or cells/molecules; (2) clinical research with interventions in humans and observations in a patient population. It is of great significance for a drug to translate the discovery in its basic research to clinical contributions and finally get approved by the Food and Drug Administration (FDA). Thus, in this study, we first examine the distribution and dynamics of basic research and clinical research in the field of drug research, to describe the overall status of drug translational research. Second, for a given drug $d$, we define the translational lag of $d$ as the time interval between the publication time of its earliest basic article ($T_{EB}$) and the publication time of its earliest clinical article ($T_{EC}$), as expressed as follows.

$$Translational\ Lag\ (d) = T_{EC} - T_{EB} + 1 \quad (3)$$

With this indicator, we analyze the distribution of translational lags for drugs that have been successfully translated into clinical science. Meanwhile, by observing the triangle of medicine for drug translational research, we conduct a preliminary investigation on the multidisciplinary collaboration between basic science and clinical science in drug research, which should be encouraged to accelerate the translation of a drug.

**Results**

*Overview of drug translational research*

**Table 1. Distribution of different types of drug-related articles in PubMed during 1965-2018.**

| Article Type | Number of articles | Percentage in drug-related articles (%) | Percentage in PubMed (%) |
|---|---|---|---|
| Animal (A) | 1,009,427 | 11.10 | 3.45 |
| Cell/Molecule (C) | 1,106,841 | 12.17 | 3.78 |
| Human (H) | 2,274,807 | 25.01 | 7.77 |
| AC | 1,892,929 | 20.82 | 6.46 |
| CH | 1,260,346 | 13.86 | 4.30 |
| AH | 266,731 | 2.93 | 0.91 |
| ACH | 677,790 | 7.45 | 2.31 |
| Other | 605,065 | 6.65 | 2.07 |
| Total | 9,093,936 | 100 | 31.05 |

As shown in Table 1, there are a total of 9,093,936 articles identified as drug-related research, accounting for 31.05% of the whole PubMed. Of these, basic science contributed nearly half of the total drug-related articles (13.69% of the whole PubMed): 1,009,427 A articles (11.10%), 1,106,841 C articles (12.17%) and 1,892,929 AC articles (20.82%); while the number of H article, which represents the clinical researches in drug research, only accounts for 25.01% of the total number of all drug-related articles. Furthermore, the collaboration between basic science and clinical science contributed to 2,204,867 articles in drug research, accounting for 24.24% of the total number of all drug articles (7.52% of the whole PubMed).

There are three collaboration types between basic science and clinical science in drug research, that is, CH, AH, and ACH. Specifically, CH with 1,260,346 articles (13.86% of the total number of all drug articles), is the most common collaboration type between basic and clinical science. Meanwhile, there is 7.45% of the total number of all drug articles (677,790) are the results of the collaborations among A, C and H scientists. However, there are only 2.93% of drug articles (266,731) were contributed by the collaboration between A and H scientists.

Figure 1 shows the changes in the number of different types of articles in drug research from 1841 to 2018, from which we can see that all seven curves exhibit a sharply increasing trend since the year 1975. For example, the number of H articles is 103,747 in 2018, which increased more than 9 times compared to that in 1975 (10,948). We can also observe that the number of AC articles ranked first place from 1975 to 1997, while after 1997 this number surpassed the number of H articles and ended in second place. For the rate of growth of the number of articles after 1970, H and C articles ranked in first and second places. Specifically, the rank for the number of H articles increased from second (1973) to first place since 1997, while the rank for the number of C articles jumped from fifth (1979) to third since 2010. In 2018, the number of CH, A, and ACH ranked fourth, fifth, and sixth places, respectively. AH with 13114 articles in 2018, has been the last place since 1970.

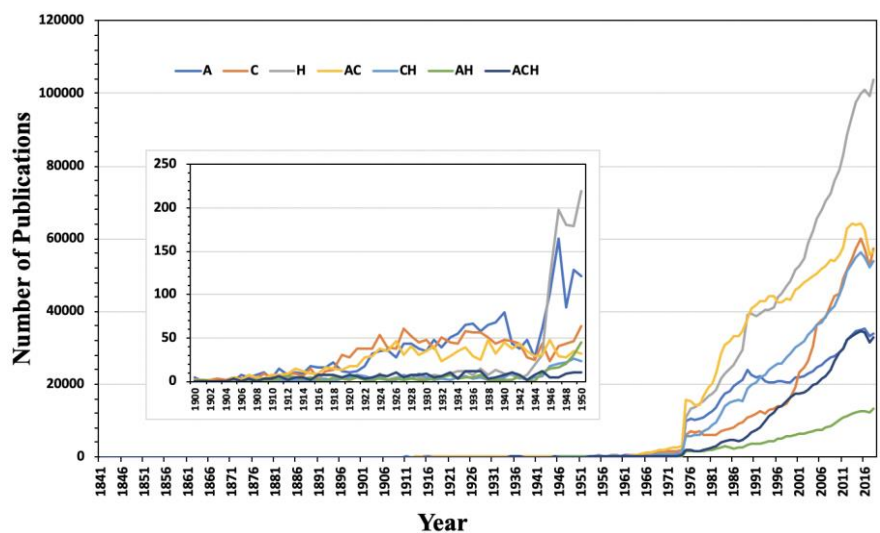

**Figure 1. Changes in the number of different types of articles in drug research from 1841 to 2018. The inset represents the changes in the number of articles in drug research between 1900-1950.**

The inset shows the changes in the number of different types of drug-related articles from the period of 1900-1950. Up to 1946, basic articles (including A, C, and AC articles) have been always in the first three places, while there are few clinical articles (H articles). The number of H articles rose from sixth place in 1942 to second place in 1945 and became first place in 1946, from less than 10 in 1990 to over 220 H articles in 1950. However, other human-related articles (including CH, AH, and ACH) have been in the last three places (less than 50) in this period.

*Translational lag analysis of drug research*

Figure 2(a) shows the changes in the annual number of non-translated drugs, drugs with their earliest basic research, and drugs with their earliest clinical research during the period of 1841-2018. We observe that the number of all three kinds of drugs exhibits increasing-decreasing trends. However, the peak values of these three curves appeared in various years, that is, 3010 non-translated drugs peaked in 2005, 1898 drugs with their earliest basic research peaked in 1975 and 1003 drugs with earliest clinical research peaked in 1975. Before 1895, there were only several drugs being studied, although every drug had been translated into clinical applications successfully. After 1895, the number of drugs studied sharply increased, and the annual number of drugs with their first basic research has been much more than that of drugs with their first clinical research. Since 1977, the number of non-translated drugs showed a sharp increase, and the other two began to decrease. Although the number of drugs beginning their basic research exhibits an evident declining trend, the number of drugs entering into their clinical research has been keeping stable. Meanwhile, the number of non-translated drugs had dramatically increased with fluctuation until 2005, then it also started to decrease.

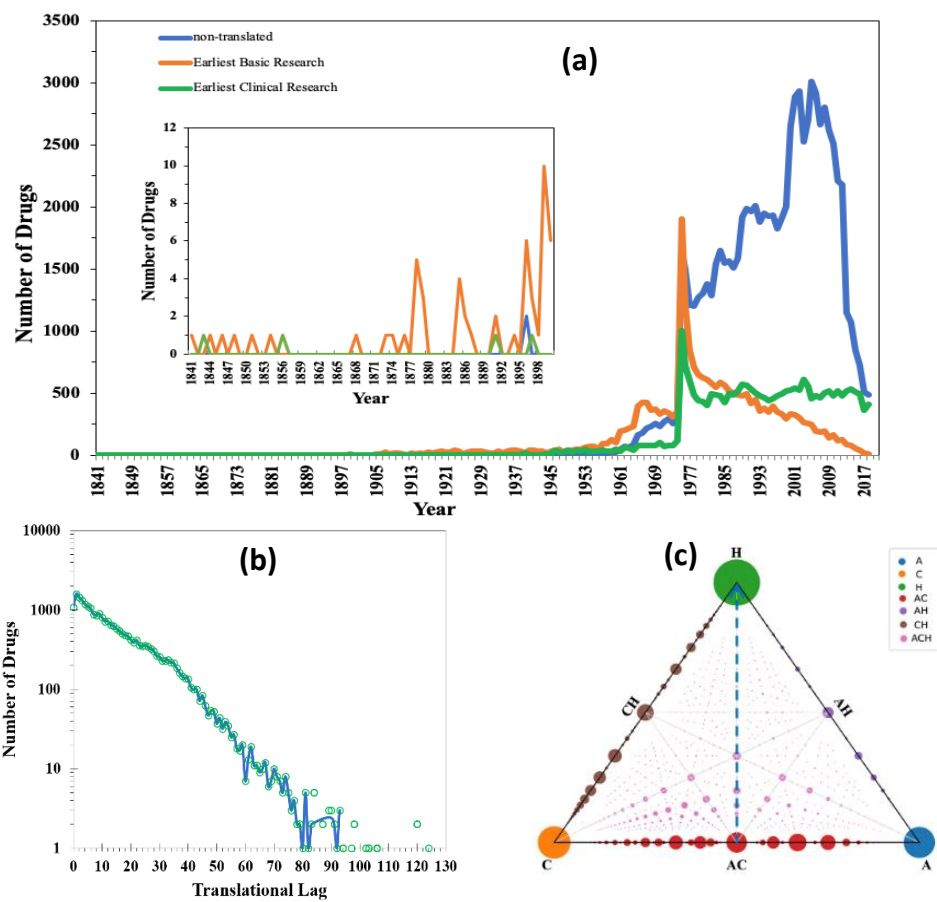

**Figure 2. Translational analysis of drug research.** (a) Changes in the annual number of non-translated drugs, drugs with their earliest basic research, and drugs with their earliest clinical research over time, respectively. (b) The distribution of translational lags in drug research. (c) Drug research mapped onto the Triangle of Biomedicine. Vertices of the triangle represent animal (A), cell/molecule (C), and human (H) research. Dotted line indicates the translational axis from basic research to clinical research. Size of the circle reflects the number of articles sharing the same coordinate.

Figure 2(b) plots the distribution of translational lag for all 24,410 drugs that have been translated successfully in our dataset, which follows an exponential distribution ($P < 10^{-4}$). This distribution reveals that the mean value of translational lag for all 24410 drugs is 14.38 years with an interquartile range of 4 to 21. Figure 2 (c) plots the different types of articles onto the triangle of medicine, in which the dotted blue line represents the translational axis and the size and density of circles display the number of articles. This picture reveals three important findings. First, the article number: H > C > A and H < A + C. Second, the number of collaborations between A and C is much more than that between H and the other two types, that is, AC > AH or CH or ACH. Finally, A and C are often dominant in ACH collaborations.

**Discussion and conclusions**

This paper examines drug research along the translational continuum using PubMed articles. The goals of this paper are to investigate the status of drug translational research, the distribution of transnational lags for all drugs as well as the multidisciplinary collaborations between basic science and clinical science in drug research. We observed that only a small percentage (18.1%) of drugs /compounds studied in PubMed have been translated successfully. Through our analysis, we found that there was a dramatic decline in the annual number of new drugs starting their first basic research since 1975, although the annual number of new drugs entering into clinical research keep the same in the same period. Meanwhile, the annual number of drugs that have not entered into clinical research exhibit a sharp increase. With these findings, we can conclude that fewer new drugs/compounds have been discovered in these years, and more and more resources might have been used for drug repurposing or modification to avoid investment risks. Moreover, we revealed that, in drug research, basic science and clinical science are less cooperative, and it seems that they are more inclined to intra-disciplinary collaboration.

There are several limitations in our paper. Because PubMed makes available only the titles and abstracts, some articles studying drugs might haven't been included. Full texts would benefit the study. On the other hand, although this paper showed the distribution of drug translational lags, it couldn't offer deeper clue as to factors affecting this time interval. In our future work, we will further explore the relationships between the drug transnational lags and team collaboration, team diversity as well as multidisciplinary collaboration.

**Acknowledgments**

This work was supported by the National Natural Science Foundation of China (grant no. 72204090). This work was also supported by the Ministry of Education of Humanities and Social Science Project (grant no. 22YJC870014). The computation was completed in the HPC Platform of Huazhong University of Science and Technology.

**References**

Hutchins, B. I., Davis, M. T., Meseroll, R. A., & Santangelo, G. M. (2019). Predicting translational progress in biomedical research. *PLOS Biology*, 17(10), e3000416.
Lee, J., Yoon, W., Kim, S., Kim, D., Kim, S., So, C. H., & Kang, J. (2019). BioBERT: A pre-trained biomedical language representation model for biomedical text mining. *Bioinformatics*, 1–7.
Lee, K., Clyne, M., Yu, W., Lu, Z., & Khoury, M. J. (2019). Tracking human genes along the translational continuum. *npj Genomic Medicine*, 4(1), 25.
Li, X., Tang, X., & Cheng, Q. (2022). Predicting the clinical citation count of biomedical papers using multilayer perceptron neural network. *Journal of Informetrics*, 16(4), 101333.
Li, X., & Tang, X. (2021). Characterizing interdisciplinarity in drug research: A translational science perspective. *Journal of Informetrics*, 15(4), 101216.
Li, X., Rousseau, J. F., Ding, Y., Song, M., & Lu, W. (2020). Understanding drug repurposing from the perspective of biomedical entities and their evolution: Bibliographic research using aspirin. *JMIR medical informatics*, 8(6), e16739.
Song, M., Kim, W. C., Lee, D., Heo, G. E., & Kang, K. Y. (2015). PKDE4J: Entity and relation extraction for public knowledge discovery. Journal of biomedical informatics, 57, 320-332.
Weber, G. M. (2013). Identifying translational science within the triangle of biomedicine. *Journal of Translational Medicine*, 11(1), 1–10.
Wei, C. H., Allot, A., Leaman, R., & Lu, Z. (2019). PubTator central: automated concept annotation for biomedical full text articles. Nucleic acids research, 47(W1), W587-W593.

Williams, R. S., Lotia, S., Holloway, A. K., & Pico, A. R. (2015). From scientific discovery to cures: Bright stars within a galaxy. *Cell*, 163(1), 21-23.

Xu, J., Kim, S., Song, M., Jeong, M., Kim, D., Kang, J., ... & Ding, Y. (2020). Building a PubMed knowledge graph. Scientific data, 7(1), 1-15.